\documentclass[sn-mathphys,Numbered]{./sn/sn-jnl}

\usepackage{graphicx}%
\usepackage{multirow}%
\usepackage{amsmath,amssymb,amsfonts}%
\usepackage{amsthm}%
\usepackage{mathrsfs}%
\usepackage[title]{appendix}%
\usepackage{xcolor}%
\usepackage{textcomp}%
\usepackage{manyfoot}%
\usepackage{booktabs}%
\usepackage{algorithm}%
\usepackage{algorithmicx}%
\usepackage{algpseudocode}%
\usepackage{listings}%
\usepackage{longtable}
\usepackage[utf8]{inputenc}
\usepackage[T1]{fontenc}

\graphicspath{{./}{figures/}}

\theoremstyle{thmstyleone}%
\theoremstyle{thmstyletwo}%

\theoremstyle{thmstylethree}%

\begin{document}

\title[]{Circumstellar Material Ejected Violently by A Massive Star Immediately before its Death}


\author[1,2,3,*]{\fnm{Jujia} \sur{Zhang}}

\author[1,2,4]{\fnm{Han} \sur{Lin}}

\author[5,6]{\fnm{Xiaofeng} \sur{Wang$^\dag$}}
\author[1,2,3]{\fnm{Zeyi} \sur{Zhao}}
\author[1]{\fnm{Liping} \sur{Li}}

\author[5]{\fnm{Jialian} \sur{Liu}}
\author[5]{\fnm{Shenyu} \sur{Yan}}
\author[5]{\fnm{Danfeng} \sur{Xiang}}
\author[7,8]{\fnm{Huijuan} \sur{Wang}}
\author[1,2,3]{\fnm{Jinming} \sur{Bai}}

\affil[1]{\orgdiv{Yunnan Observatories (YNAO)}, \orgname{Chinese Academy of Sciences (CAS)}, \city{Kunming}, \postcode{650216}, \country{China}}

\affil[2]{\orgdiv{Key Laboratory for the Structure and Evolution of Celestial Objects}, \orgname{CAS}, \city{Kunming}, \postcode{650216}, \country{China}}

\affil[3]{\orgdiv{International Centre of Supernovae}, \orgname{Yunnan Key Laboratory}, \city{Kunming}, \postcode{650216}, \country{China}}

\affil[4]{\orgdiv{Key Laboratory of Radio Astronomy and Technology}, \orgname{CAS}, \city{A20 Datun Road, Chaoyang District, Beijing}, \postcode{100101}, \country{China}}

\affil[5]{\orgdiv{Department of Physics}, \orgname{Tsinghua University}, \city{Beijing}, \postcode{100084}, \country{China}}

\affil[6]{\orgdiv{Beijing Planetarium}, \orgname{Beijing Academy of Science and Technology}, \city{Beijing}, \postcode{100044}, \country{China}}

\affil[7]{\orgdiv{National Astronomical Observatories}, \orgname{CAS}, \city{Beijing}, \postcode{100101}, \country{China}}

\affil[8]{\orgname{School of Astronomy and Space Science}, \orgname{University of Chinese Academy of Sciences}, \city{Beijing}, \postcode{101408}, \country{China}}

\affil[*]{jujia@ynao.ac.cn}

\affil[\dag]{wang\_xf@mail.tsinghua.edu.cn}

\abstract{Type II supernovae represent the most common stellar explosions in the Universe, for which the final stage evolution of their hydrogen-rich massive progenitors towards core-collapse explosion are elusive. The recent explosion of SN 2023ixf in a very nearby galaxy, Messier 101, provides a rare opportunity to explore this longstanding issue. With the timely high-cadence flash spectra taken within 1-5 days after the explosion, we can put stringent constraints on the properties of the surrounding circumstellar material around this supernova. Based on the rapid fading of the narrow emission lines and luminosity/profile of $\rm H\alpha$ emission at very early times, we estimate that the progenitor of SN 2023ixf lost material at a mass-loss rate $\dot{\rm M} \approx 6 \times 10^{-4}\, \rm M_{\odot}\,a^{-1}$ over the last 2-3 years before explosion. This close-by material, moving at a velocity $v_{\rm w} \approx 55\rm \, km\,s^{-1}$, accumulates a compact CSM shell at the radius smaller than $7 \times 10^{14}$ cm from the progenitor. Given the high mass-loss rate and relatively large wind velocity presented here, together with the pre-explosion observations made about two decades ago, the progenitor of SN 2023ixf could be a short-lived yellow hypergiant that evolved from a red supergiant shortly before the explosion. }


\maketitle
\section{Introduction} 
\label{sec:intro}
The mass loss history of massive stars during their evolution towards core-collapse supernovae (CCSNe) is unclear \cite{2017hsn..book..403S}. The final-stage evolution and the resultant circumstellar environments have led to a rich diversity of such explosions. To establish a link between the explosion of CCSNe and the late-time evolution of massive stars, it is necessary to capture the first-light signals of the SN explosions, i.e., the flashed spectra \citep{2014Natur.509..471G}, due to ionization of the circumstellar material (CSM)/stellar wind by ultraviolet/high energy photons from shock breakout cooling. In particular, the evolution of such flashed spectral features (i.e., narrow emission lines) can help constrain the density profile and hence the structure of the CSM around the SN. 
 
As the flashed emission features superimposed on the continuum last only a few days after explosions for regular hydrogen-rich CCSNe (known as Type II supernovae, SNe II), it is not easy to capture their complete evolution. The first observed case of SN II with a flash spectrum varying on timescales of hours is SN 2013fs \citep{2017NatPh..13..510Y}, which suggests a dense CSM enclosing the SN. However, the scarcity of such high-quality spectroscopic observations hinders our understanding of this important issue.  

SN 2023ixf is such an event that recently exploded in the nearby spiral galaxy M101 at a distance D = 6.85 Mpc \citep{2022ApJ...934L...7R}, providing an opportunity to realize the properties of its progenitor and the surrounding environment through high cadence spectroscopic observations at extremely early phases. Unlike distant objects, an SN exploding at D $<\, \sim$ 10 Mpc allows a variety of unusual observational techniques to achieve higher spatial, temporal, and wavelength resolution.

Itagaki first reported the discovery of SN 2023ixf on May 19.73, 2023 \cite{2023TNSTR1158....1I} (UT dates are used throughout this paper). Given the appearance of prominent emission features in the spectrum obtained within a few hours of discovery, this transient was classified as an SN II that interacted with the nearby circumstellar material (CSM)\cite{2023TNSAN.119....1P}. SN 2023ixf was detected in pre-discovery images about one day before the first report. These give a reliable limit on the explosion date, e.g., May 18.76 (MJD = 60082.76) derived from modeling of the early light curve (Li Gaici et al. in review). The first detection occurred a few hours later than this date.

Given the pre-explosion detection \cite{2023Xiangprep}, early-time discovery, and rapid follow-up, SN 2023ixf is an ideal case to map the observed properties of a CCSN to its progenitor. As a result, a worldwide observational campaign for this SN was initiated, including the whole-band electromagnetic radiation \cite{2023arXiv230604827G,2023arXiv230604721J,2023TNSAN.180....1M,2023arXiv230616397M,2023arXiv230607964S} and neutrino \cite{2023arXiv230614717G} observations. 

In this paper, we present detailed analysis and constraints on the nearby compact CSM around SN 2023ixf based on our efforts in high-cadence (i.e., $\sim$ one observation per 2 - 3 hours on May 20) follow-up observations of the flash spectra in intermediate-resolution mode. It is the first report of SN 2023ixf at such a high cadence follow-up with spectral resolution above 3000. The configuration of the nearby CSM is crucial to reveal the mass-loss history and properties of the progenitor.

\begin{figure}
\centering
\includegraphics[width=12cm,angle=0]{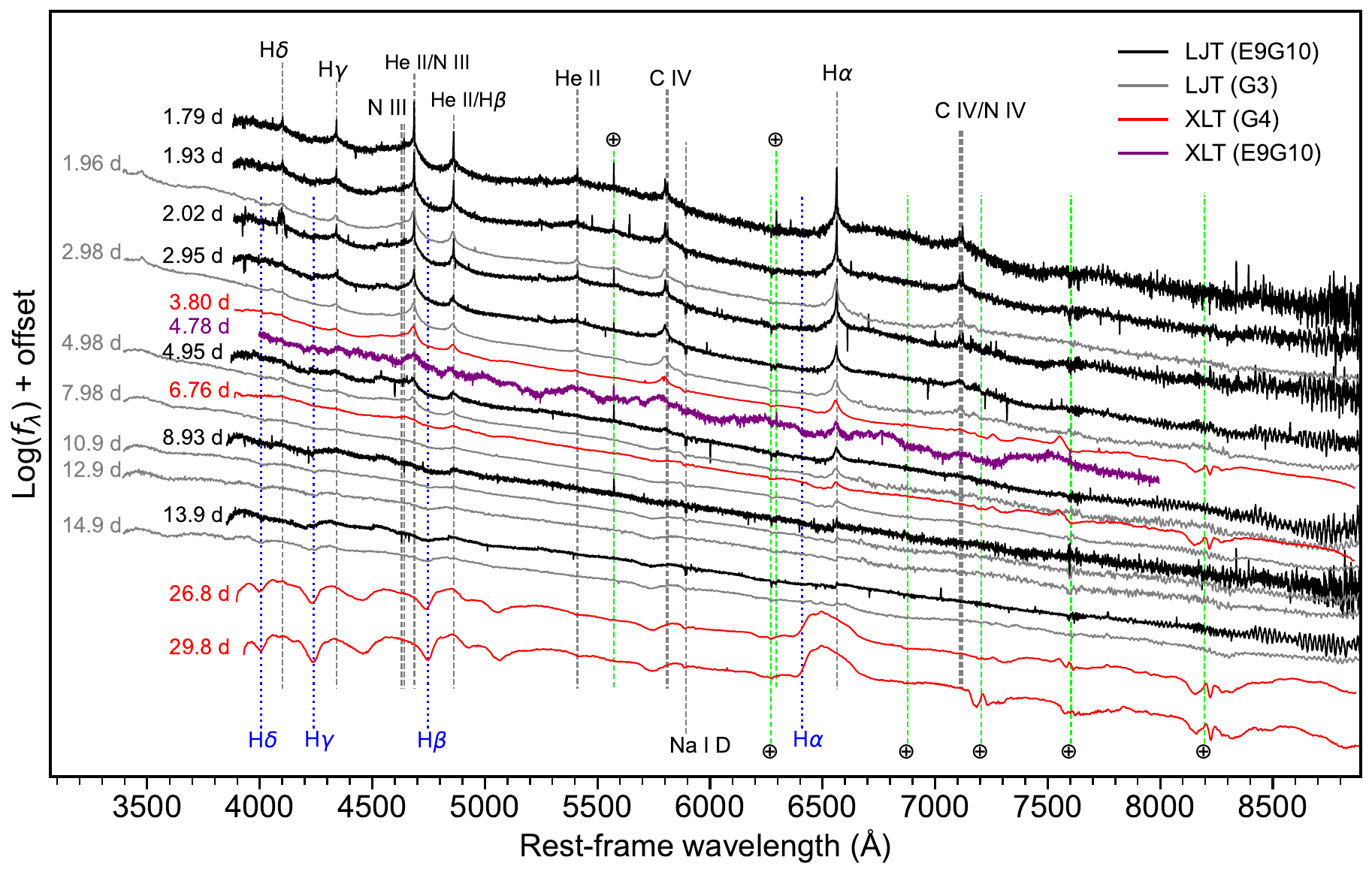}
 \caption{Spectra of SN 2023ixf obtained by LJT and XLT. The dashed grey lines mark spectral features at rest wavelength, and the dotted blue lines at a velocity of -7000 km s$^{-1}$. Residuals from incomplete removal of telluric absorptions and skyline emissions are marked as green dashed-dotted lines with an Earth symbol. The observation journal is listed in Table  \ref{Tab:Spec_log} . }
\label{<spe_whole>}
\end{figure}

\begin{figure}
\centering
\includegraphics[width=12cm,angle=0]{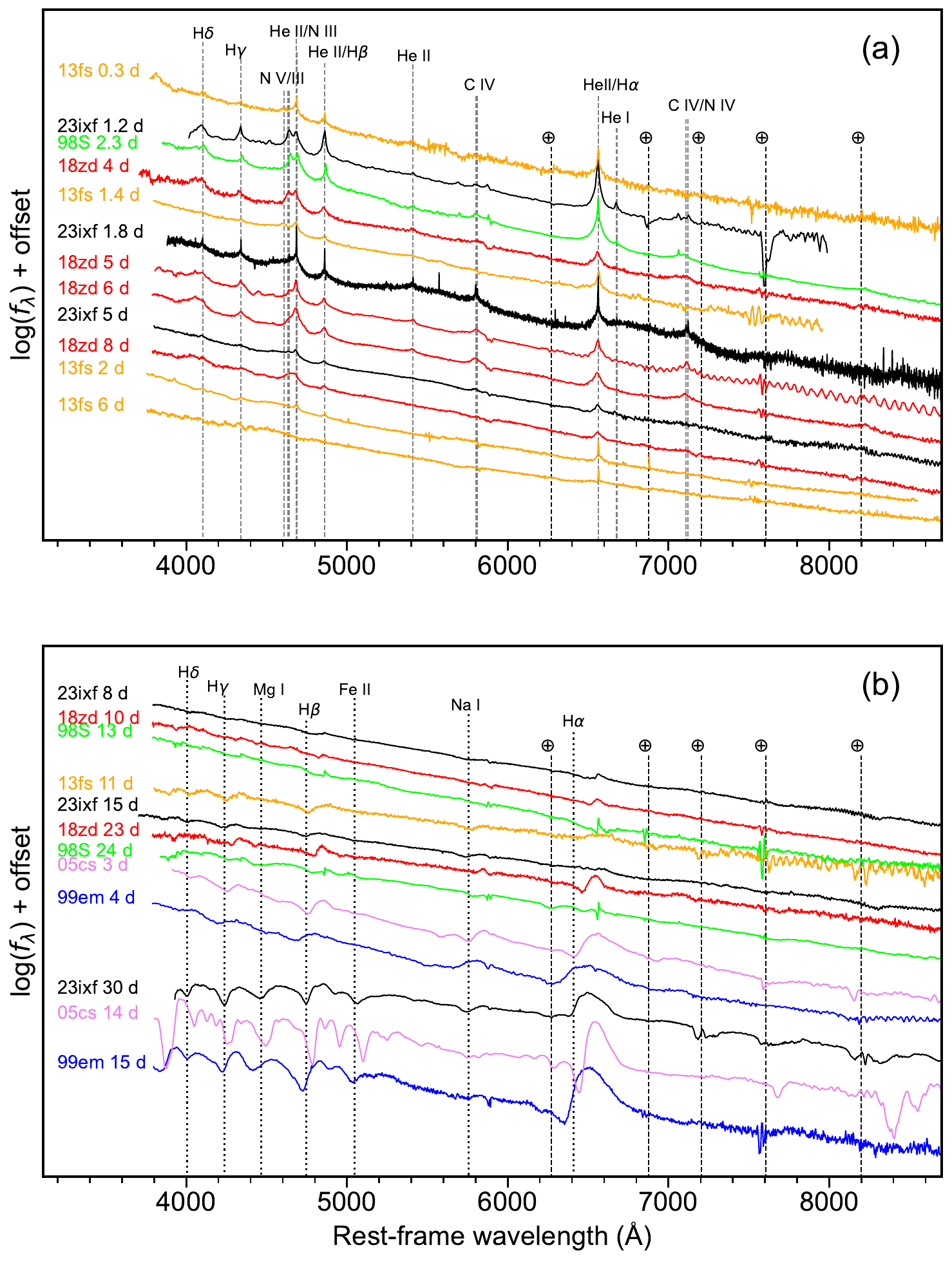}
\caption{Spectral comparison of SN 2023ixf at two phases. (a) Flash-ionisation-phase spectra of SN 2023ixf compared with spectra of SN 1998S \citep{2000ApJ...536..239L,2001MNRAS.325..907F,2015ApJ...806..213S}, SN 2013fs \citep{2017NatPh..13..510Y}, and SN 2018zd \citep{2020MNRAS.498...84Z}. (b) Spectra of SN 2023ixf spanning from the first week to the first month, along with those of SN 1998S, SN 1999em\citep{2001ApJ...558..615H,2002PASP..114...35L}, SN 2005cs \citep{2009MNRAS.394.2266P}, and SN 2013fs. The spectrum at $t\approx 1.2$d obtained by Liverpool telescope \citep{2023TNSAN.119....1P} was plotted.}
\label{<spe_comp>}
\end{figure}

\section{Spectral Observations}
We quickly responded to SN 2023ixf on 2023 May 20. This SN was monitored in middle-resolution spectra at Li-Jiang 2.4-m telescope (LJT; \citealp{2015RAA....15..918F}) with YFOSC (Yunnan Faint Object Spectrograph and Camera; \citealp{2019RAA....19..149W}) cross-dispersion model with Grism-10 (G10) and Echelle-9 (E9) three times over the course of this day. Using the slit at a width of 0".8, the resolution of G10E9 reaches around 3500. We also obtained several high- and low-resolution spectra with LJT and Xing-Long 2.16m telescope (XLT) in the following month, as presented in Fig. \ref{<spe_whole>}.

The equivalent width (EW) of interstellar Na~{\sc i}  D1 and D2 lines in two high-resolution spectra of LJT (i.e., Fig. \ref{<spe_NaID>} online) are EW$_{D1} = 0.11 \pm 0.01$ \AA\, and EW$_{D2} = 0.18 \pm 0.01$ \AA, which produces the extinction of the host galaxy $E(B-V)_{\rm host} = 0.031 \pm 0.001$\, mag via the empirical relation \cite{2012MNRAS.426.1465P}. Combining the Galactic reddening $E(B-V)_{\rm MW} = 0.008$\, mag measured by \citet{2011ApJ...737..103S}, the total extinction of SN 2023ixf adopted in this paper is $E(B-V)_{\rm total} = 0.04$\, mag.

The narrow emission lines of hydrogen dominate the early spectra of SN 2023ixf, accompanied by the emissions of higher ionized elements, e.g., He~{\sc ii}, N~{\sc iii}, and C~{\sc iv}, which are usually called flashed-ionized features \citep{2014Natur.509..471G,2016ApJ...818....3K}. These flash features are generated by the surrounding wind material, which was ionized by X-rays and ultra-violet photons from the shocked ejecta.

SN 2023ixf shows He~{\sc i} $\lambda$6678 emission in the classification spectrum at $t \approx 1.2$ d, as seen in Fig. \ref{<spe_comp>} (a). This line disappeared half a day later and appeared again at $t\sim 5$ d, which implies the process of ionization and recombination of helium due to the rise and fall of temperature. In this panel, the flash features in SN 2023ixf appear to evolve more slowly than in SN 2013fs but faster than in SN 2018zd and SN 1998S. The duration of these interaction signatures relates to the mass-loss rate and the extensibility of surrounding material. 

About one week after the explosion, SN 2023ixf started to enter the photospheric phase when the flash spectral features nearly vanished, and the P-cygni profile of the H Balmer series became visible. The relatively weak Balmer lines and blue continuum are likely due to continuous heating by SN-CSM interaction. SN 2023ixf evolves the clear spectral lines of typical SNe II about one month after the explosion. The spectrum of this SN at t$\sim$ 30 d is similar to that of the normal SN II 1999em at t $\sim$ 3 d and low luminous SN II 2005cs at t $\sim$ 4 d, as seen in Fig. \ref{<spe_comp>} (b). At this phase, the emission of broad H$\alpha$ shows significant asymmetry with the suppressed red part due to the occultation of high opaque \citep{2001MNRAS.326.1448C}.

\section{Analysis of the early flash spectra}

\begin{figure*}
\centering
\includegraphics[width=12cm,angle=0]{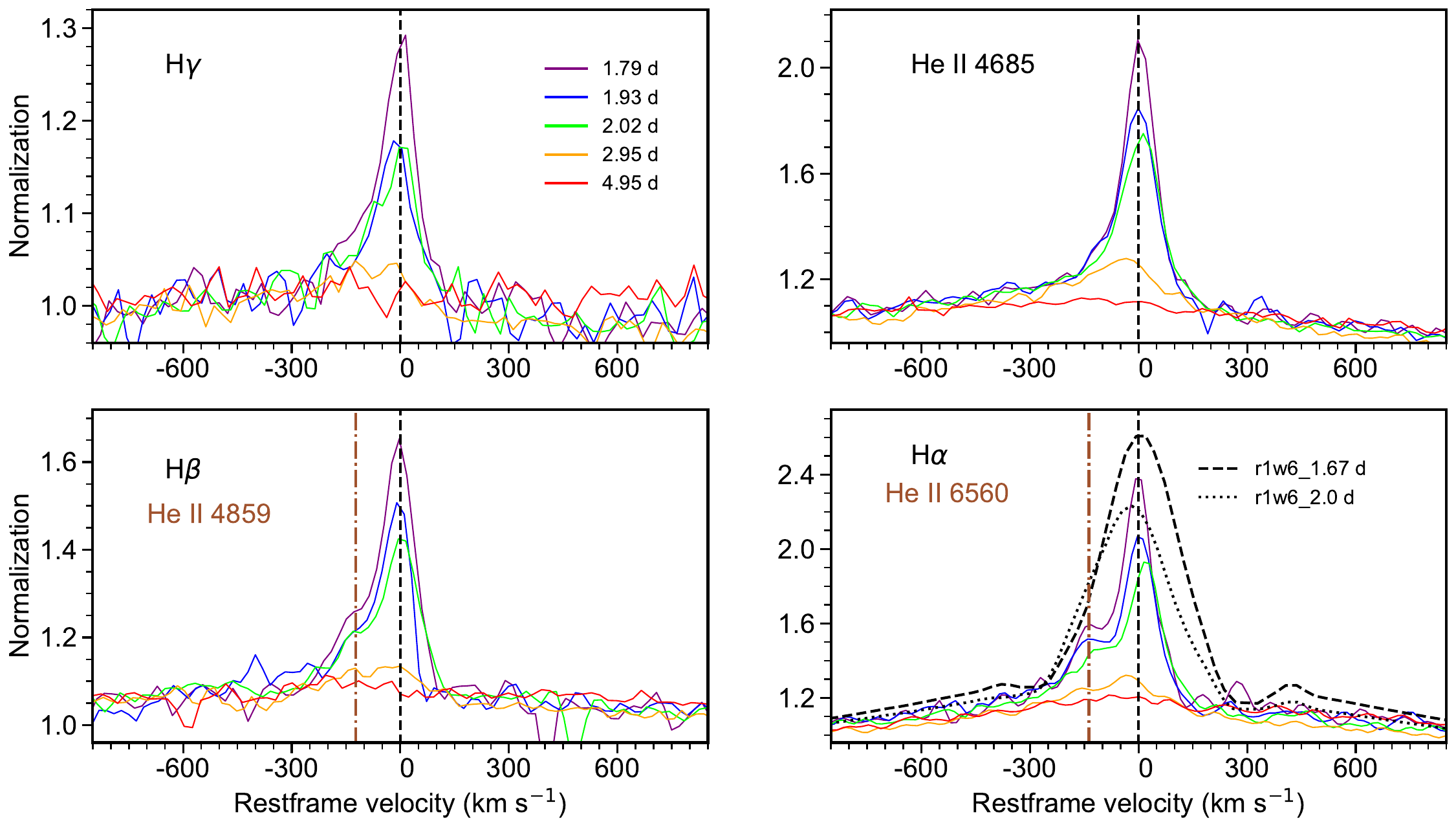}
 \caption{Flash emissions of SN 2023ixf at the first five days. The black dashed lines marked the primary spectral line of each panel at rest, and the sienna dash-dotted lines marked the secondary spectral line of this panel at rest. The $H\alpha$ emission in two r1w6 spectra are plotted. }
\label{<spe_flash>}
\end{figure*}

In the early phase, while the SN photosphere is in the pre-shock ionized CSM, the flash emissions are generated by recombining the CSM out of the photosphere and characterized by a narrow Gaussian core superimposed on a broad Lorentzian wing. The broad wings of these emissions are formed by electron scattering. The narrow component, originating from the unscattered light, evolves on timescales of hours at $t < 2$ d (as seen in Fig. \ref{<spe_flash>}). These rapid variations allow us to investigate the properties of the CSM.

A P-cygni profile with narrow emission due to CSM expansion at a velocity of $\sim$100 $\rm km\, s^{-1}$ can explain the prominent double-peaked profiles of H$\alpha$ and H$\beta$ at $t < 2$d. However, we prefer the contamination of He~{\sc ii}$\lambda$6560 and 4859 over the P-cygni profile due to the simultaneous detection of He~{\sc ii}$\lambda$4685. He~{\sc ii}$\lambda$6560 and 4859 are usually unresolved in most previous observations of SNe due to expansion effects and low spectral resolution, which results in an overestimate of the hydrogen line flux. To remove the contamination of He~{\sc ii}, we use a Lorentz profile and two Gaussian profiles to fit the spectra around the H$\alpha$ emission (as seen in Fig. \ref{<ha_3fit>} online). 

The observed spectral profile is the convolution of the true spectral line and the instrumental profile. The true FWHM of spectral line can be derived through FWHM = (FWHM$^2_{\rm obs}$ $-$ FWHM$^2_{\rm inst})^{1/2}$, where instrumental expansion estimated by the skylines is  $85\pm 3\, \rm{km}\ s^{-1}$.

The corrected FWHM of the narrow H$\alpha$ emissions broadens from $\sim55 \rm km\, s^{-1}$ to $\sim80\, \rm km\, s^{-1}$ on the timescale of hours at $t <\sim 2$ d, shown in Table \ref{Tab:ha} . Conversely, the FWHM of the broad component remains consistent at $\sim\, 2500\, \rm km\, s^{-1}$. The narrow component fades sharply a day later, and the profile becomes considerably wider, e.g., FWHM $\approx 140\,\rm km s^{-1}$. An intermediate component replaces this narrow component with an FWHM larger than 800 $\rm km s^{-1}$ at $t\approx 5$ d. The evolution of the narrow lines is less pronounced than the above in our low-resolution spectra at the same phase. Thus, it becomes difficult to understand the nature of the flash feature in SN 2023ixf if the spectroscopic observations start at t $> \sim$ 2 d and the spectral resolution is not high enough.

As shown in the model spectra \cite{2001MNRAS.326.1448C}, the width of the H$\alpha$ is positively correlated with the Thomson optical depth $\tau_{T}$ and electron temperature $T_e$. The temperature derived from the blackbody increases by only 3\% in the first three spectra and may not significantly affect the spectral lines.  Besides, the ratio ($U$) of the narrow component to the total $H\alpha$ line has a relation to $\tau_{T}$ (i.e., $U \approx [1-exp(-\tau_{T})]/\tau_{T}$, \citealp{2001MNRAS.326.1448C}). In SN 2023ixf, $U(\tau_T)$ become smaller from $\sim$ 0.12 to $\sim$ 0.08, spanning from  1.79 d to 2.02 d, indicating an obvious increase in $\tau_{T}$ from $\sim 8$ to $\sim 13$. Thus, the broadening and fading seen in the narrow $H\alpha$ line could be due to the increased optical depth, suggesting a complex CSM structure enveloping this SN.

\begin{figure}
\centering
\includegraphics[width=12cm,angle=0]{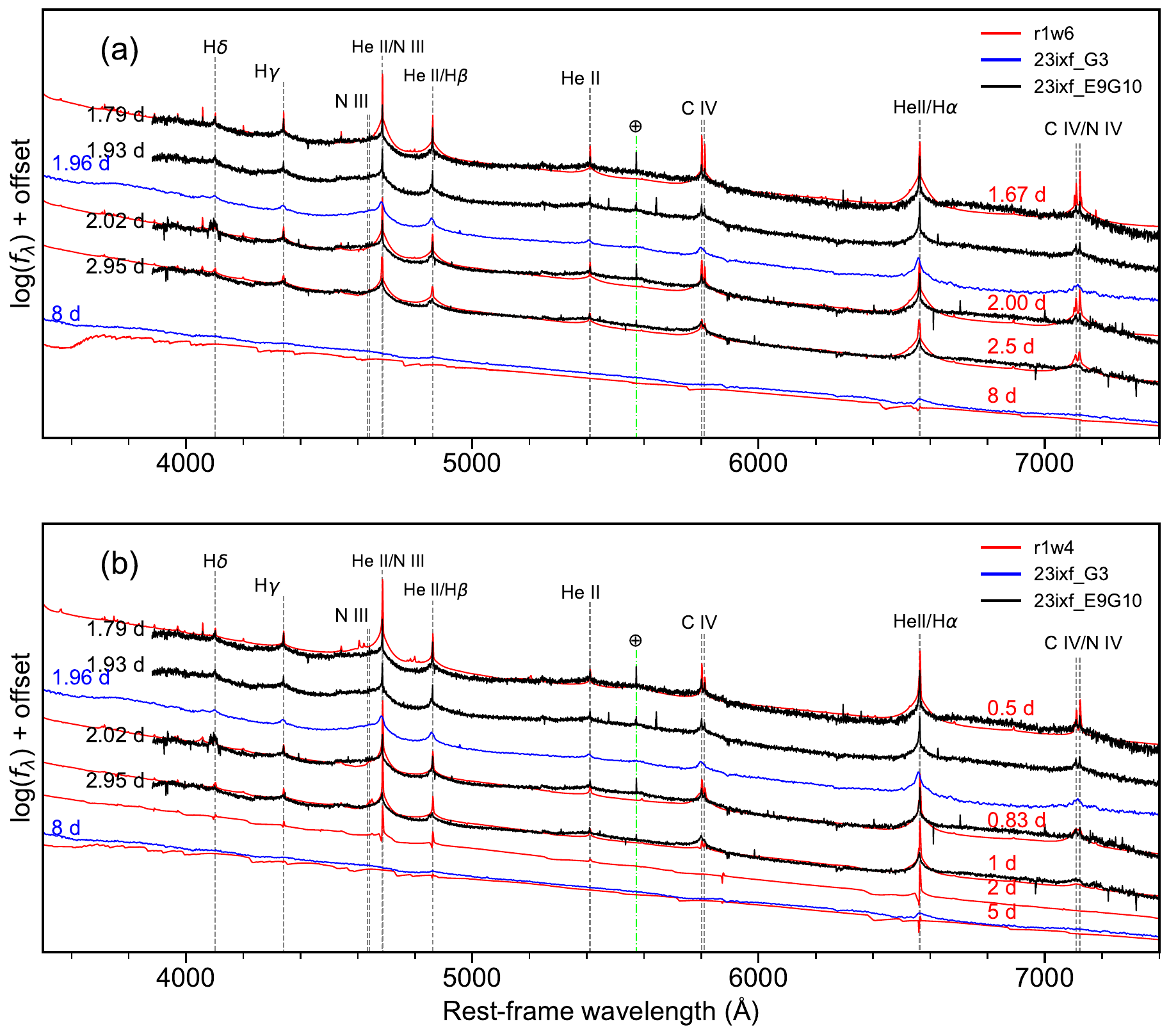}
 \caption{Spectra of SN 2023ixf at selected phases comparing with the r1w6 and r1w4 model spectra of D17. The dashed grey lines mark spectral features at rest. Residuals from incomplete removal of skyline emission (O I $\lambda$5577) is marked as a green dashed-dotted line with an Earth symbol.  }
\label{<spespe_earl_r1w6>}
\end{figure}

To understand the ionization state of SN 2023ixf, we compare observed spectra to the radiation hydrodynamics and nonlocal-thermodynamic-equilibrium (NLTE) radiative transfer model \cite{2017A&A...605A..83D} (D17). This model can produce observational signatures of SNe II with short-lived flash spectra, taking into account the different physical states of the progenitor before the explosion. Of a series of parameters distinguishing different models, only two sets can produce model spectra containing all kinds of narrow emission lines as seen in SN 2023ixf.

In Fig.\ref{<spespe_earl_r1w6>} (a), the r1w6 model of D17 closely follows the ionization features seen in SN 2023ixf at a similar phase. In r1w6 model, the progenitor star radius is $R_{\star}=501\,{\rm R}_{\odot}$, the wind mass-loss rate is $\dot{M}=1.0 \times10^{-2}\,{\rm M}_{\odot}\,{\rm a}^{-1}$, the density of CSM is $\rho_{CSM} = 1.0 \times 10^{-12} {\rm g\,cm}^{-3} $, and the radius of CSM is $R_{CSM} = 5 \times 10^{14} {\rm cm}$. 

The first four E9G10 spectra of SN 2023ixf match each spectral line shown in the model. The narrow emission in the observed spectra of SN 2023ixf and the r1w6 model decreases around 2-3 days after the explosion. 

We note that the narrow emission in r1w6 is stronger than that in SN 2023ixf, both in the depth and width of the line, e.g., H$\alpha$ in Fig. \ref{<spe_flash>}. A wider line profile in the D17 model may suggest a larger wind velocity and mass-loss rate. In addition, the spectrum of SN 2023ixf at t $\approx$ 3 d shows weaker high-ionization lines of C~{\sc iv} and N~{\sc iii} compared to the spectrum of r1w6 at $t = 2.5$ d. That could indicate a faster ionization evolution of SN 2023ixf.

The r1w4 model with a lower mass-loss rate ($\dot{M}=1.0 \times10^{-3}\,{\rm M}_{\odot}\,{\rm a}^{-1}$) in D17 can also produce the flash spectra seen in SN 2023ixf. However, this model evolves faster than the observation, as Fig. \ref{<spespe_earl_r1w6>} (b) shows. Considering the duration and intensity of the flash feature, the mass-loss rate of SN 2023ixf likely lies between the values adopted in r1w4 and r1w6. We caution that there is no difference in other progenitor parameters of these two models, such as mass and metallicity, essential to yielding observational diversity. Thus, it is best to consider the mass-loss rates of these models as a helpful reference. 

About eight days after the explosion, SN 2023ixf and r1w6 become virtually featureless, which means that the photosphere breaks through the CSM shell, and the optical depth at this phase is too high to produce spectral lines seen in the regular SN II, e.g., the P-cygni profile of neutral hydrogen.

\section{CSM Structure and Progenitor}
The duration and strength of the flash lines depend on the radius and density of the CSM and give a clue to derive the mass-loss rate of the progenitor. In Fig. \ref{<spe_flash>}, the broad emission component presents longer than the narrow one. It disappears about eight days after the explosion when the SN ejecta overtakes the outer CSM. Assuming an ejecta velocity of $10^{4}\,\rm{km}\,\rm s^{-1}$, the CSM should be confined within $7 \times 10^{14}$ cm that satisfies the radius derived by bolometric light curve (i.e., $7.4 \times 10^{14}$ cm in Fig. \ref{<bolo>} online) and the initial radius ($5 \times 10^{14}$ cm) in r1w4 and r1w6 models. A wind at a velocity $v_{\rm w}\approx 70\,\rm km\ s^{-1}$ (adopt the average velocity at $t < \sim 2$ d) can reach this distance in three years. 

The $\rm H\alpha$ luminosity $L_{\rm H\alpha}$ generated by the recombination can be used to estimate an order of magnitude of mass-loss rate via the method, i.e., $L_{\rm{H}\alpha}\,\approx\,2\,\times\,10^{39}\,\dot{\rm M}^2_{0.01}\,v^{-2}_{\rm{w},500}\,\beta\,\rm{r^{-1}_{15}\,erg\,s^{-1}}$ \cite{2013ApJ...768...47O}. Given that the recombination of the CSM produces the narrow Gaussian core and the broad Lorentzian wing, we estimate the mass-loss rate of SN 2023ixf by taking the luminosity of these two components together.

Table \ref{Tab:ha} lists the results based on the first five middle-resolution spectra. Note that the luminosity of the narrow component $L^{\rm N}_{\rm H\alpha}$ is about one-tenth of the total luminosity, so the mass-loss rate derived from the narrow component alone will be one order lower than that in this table.

The method of \cite{2013ApJ...768...47O} requires the outer radius to be twice as large as the inner since the emitting region of H$\alpha$ is from the photospheric radius to the outer radius of CSM, i.e., $\beta \ge 1$, where $\beta \equiv (r_1-r)/r$, r1, and r are the outer and inner radius of CSM. Only the spectra at $t<\sim 2$ d can satisfy this requirement due to the fast-fading interaction signal.

\begin{figure}
\centering
\includegraphics[width=12cm,angle=0]{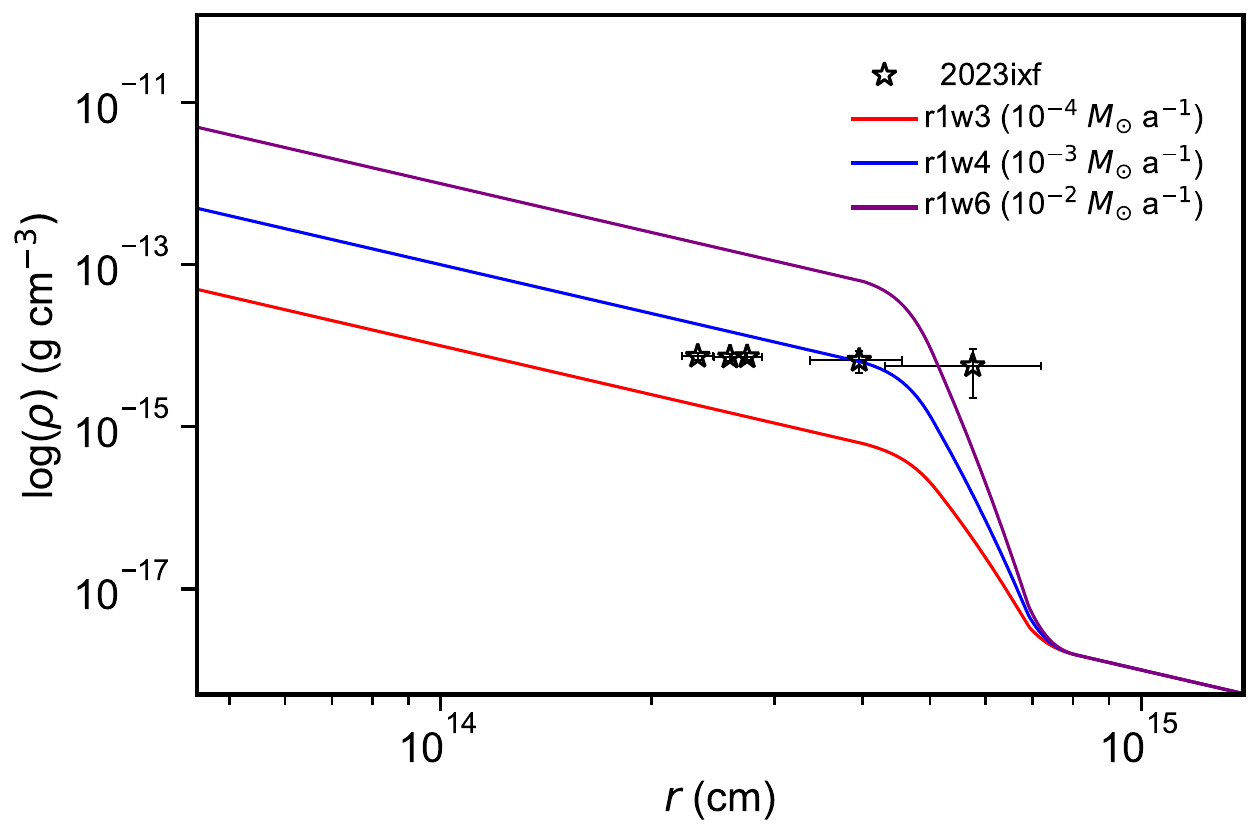}
\caption{CSM configuration around SN 2023ixf derived by $L_{\rm H\alpha}$ at $t < 5$ d. Compared with the initial density structure used for models r1w3, r1w4, and r1w6 in D17. Note the last two points of SN 2023ixf are unreliable due to their radius parameter $\beta<1$.  }
\label{<dense>}
\end{figure}

Considering the effect of supernova radiation to the environment, the result derived by the first spectrum can make a better constrain to the CSM, where the mass-loss rate is $\dot{\rm M} \approx 6 \times 10^{-4} \rm M_{\odot} a^{-1}$. Given a wind velocity $v_{\rm wind}\approx 55\, \rm km\, s^{-1}$, the wind parameter ($\omega = \dot{M} / v_{\rm w}$) of the CSM is $\sim 6 \times 10^{15} \rm g\ cm^{-1}$, which is close to that of SN 2013fs.

The mass-loss rate derived from $L_{H\alpha}$ is consistent with the early spectral modeling and X-ray observation. For example, the density evolution of SN 2023ixf is located at the side close to the r1w4 model compared to the r1w3 model in the CSM configuration of Fig. \ref{<dense>}.  Considering the flash spectral reproduced by r1w4 and r1w6 models, the spectral evolution trends, and the CSM structure, the mass-loss rate of SN 2023ixf can be on the order of $10^{-3}{\rm M}_{\odot}\,{\rm a}^{-1}$. In addition,  the mass-loss rate of SN 2023ixf estimated from X-ray detection is $\dot{M}=3.0 \times10^{-4}\,{\rm M}_{\odot}\,{\rm a}^{-1}$ for an assumed wind velocity of $v_w = 50\,\rm \,km\,s^{-1}$ \cite{2023arXiv230604827G}.

We note that estimation of the mass-loss rate varies significantly among different methods. The spectral models of r1w6 and r1w4 suggest that SN 2023ixf has a mass-loss rate ranging from $10^{-3}\rm\, to\, 10^{-2}  M_{\odot} a^{-1}$, while the recombination luminosity of $L_{H\alpha}$ produces moderate values at (6-8)$\times 10^{-4} \rm M_{\odot} a^{-1}$. The luminosity of the narrow component alone yields results in the range of (6-9)$\times 10^{-5} \rm M_{\odot} a^{-1}$. Although the estimation based on $L_{H\alpha}$ is adopted in the discussions, we do not exclude the results of other measurements, which are valuable to understand the mass loss history of massive stars.

\begin{figure}
\centering
\includegraphics[width=12cm,angle=0]{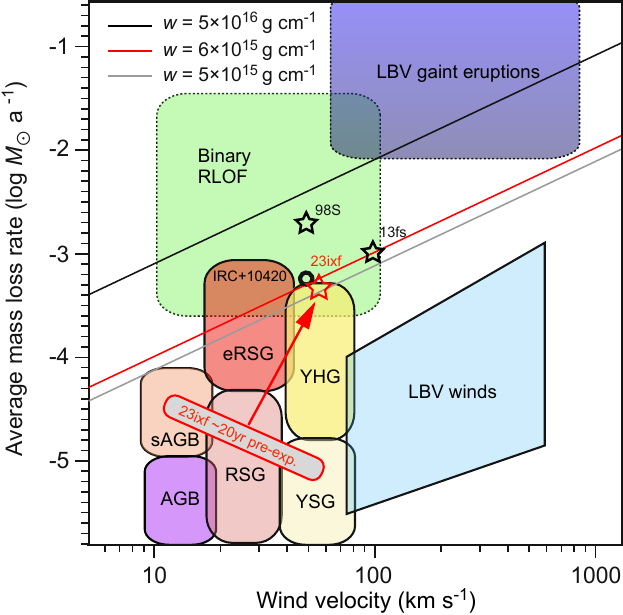}
\caption{The mass-loss rate and wind velocity of SN 2023ixf is indicated by the red star, overplotted with those of SN 1998S and SN 2013fs. The red-grey box presents the parameters for the progenitor of SN 2023ixf at $\sim$20 years before the explosion \cite{2023Xiangprep}. The red arrow indicates the change in parameter space, as required to evolve from the progenitor two decades ago to the presupernova star immediately before the explosion. 
The colored regions correspond to values for various evolved massive stars from \cite{2014ARA&A..52..487S,2017hsn..book..403S}, including the AGB, super-AGB, RSG, extreme RSG (eRSG), yellow supergiants (YSG), YHG, luminous blue variable (LBV) winds, LBV giant eruptions, and binary Roche-lobe overflow (RLOF). IRC+10420 with a well-determined high mass-loss rate is shown with circles.  The solid lines present the wind parameter $w$ related to the producing ionized spectral features. }
\label{<wind>}
\end{figure}

Finally, we attempt to map our observation to the progenitor system. The early spectral observation suggests the progenitor of SN 2023ixf has lost $\sim 2\times 10^{-3} \rm M_{\odot}$ of material over the past 2-3 years at a mass-loss rate $\dot{\rm M} \approx 6 \times 10^{-4} \rm M_{\odot} a^{-1}$ producing a compact CSM shell at a radius smaller than $7 \times 10^{14}$ cm. 

The archive images from about two decades before the explosion suggest that the progenitor of SN 2023ixf might be a dusty red supergiant (RSG) or super asymptotic giant branch (sAGB) \cite{2023Xiangprep,pledger2023possible,mayya2023identification,soraisam2023detection}. At this phase, the mass-loss rate of progenitor derived from the dust and stellar wind \cite{2023Xiangprep} is 1-2 orders lower than our estimates, as seen in Fig. \ref{<wind>}. A mechanism that enhances the mass loss 2-3 years before the explosion is required to produce the flash spectral features of SN 2023ixf. The RSG that survives such an enhanced mass-loss phase can evolve into a yellow hypergiant (YHG) by moving towards warmer temperatures and bluer colors at the end of its lifetime \citep{2014ARA&A..52..487S}.YHG is a massive star in the post-RSG phase with a high mass-loss rate, which may explode as an SN II with some significant spectral signals of SN-CSM interaction \citep{2017hsn..book..403S}.

In Fig. \ref{<wind>}, SN 2023ixf is located in the top region of the YHG in mass-loss rate vs. wind velocity space. Neighboring YHG IRC+10420, with similar mass-loss rates and wind parameters, is the prototypical post-RSG observed to evolve rapidly blue-ward during the strong mass-loss phase \citep{humphreys1997hst}. Observation of IRC+10420 provides further evidence for the RSG to YHG evolutionary channel \citep{2014ARA&A..52..487S}.

Considering the properties of CSM and mass loss history, we hypothesize that the progenitor of SN 2023ixf could have evolved from an RSG to be a short-lived YHG 2-3 years before the explosion. This assumption is tentative, and we expect to see more evidence of the progenitor system in the later observation.

\bmhead{Acknowledgments}

This work is supported by the National Key R\&D Program of China with No. 2021YFA1600404, the National Natural Science Foundation of China (12173082), and science research grants from the China Manned Space Project with No. CMS-CSST-2021-A12, the Yunnan Province Foundation (202201AT070069), the Top-notch Young Talents Program of Yunnan Province, the Light of West China Program provided by the Chinese Academy of Sciences, the International Centre of Supernovae, Yunnan Key Laboratory (No. 202302AN360001). X.Wang is supported by the National Natural Science Foundation of China (NSFC grants 12288102, 12033003, and 11633002), the Scholar Program of Beijing Academy of Science and Technology (DZ:BS202002), and the Tencent Xplorer Prize. 

We acknowledge the support of the staff of the LJT and XLT. Funding for the LJT has been provided by the CAS and the People's Government of Yunnan Province. The LJT is jointly operated and administrated by YNAO and the Center for Astronomical Mega-Science, CAS.

\bmhead{Author contributions}
Jujia Zhang initiated the study, conducted analysis and wrote the manuscript. Han Lin analysed the flash spectra and calculated the mass loss rate. Xiaofeng Wang and Jinming Bai triggered rapid follow-up and designed the observation.  Zeyi Zhao analysed the light curves and calculated the bolometric luminosity. Liping Li, Jialian Liu, Shenyu Yan, Danfeng Xiang and Huijuan Wang contribute to observation and data reduction.

\clearpage 
\bibliography{SN2023ixf}

\clearpage 

\begin{appendices}

\section{}~\label{sec:Appendix}
Supplemental materials, including tables and figures of the observational journal, calculation results,  spectral lines, and bolometric light curves.


\setcounter{table}{0} 
\renewcommand{\thetable}{S\arabic{table}}
\begin{table*}[!th]
\caption{Journal of spectroscopic observations of SN 2023ixf}
\scriptsize
\begin{tabular}{lccccc}
\hline\hline
Date (UT) & MJD & Epoch (d)$^a$ & Range (\AA) & Spec. Res.  & Telescope+Inst.\\
May 20, 2023	&	60084.55	&	1.79	&	3850-9160	&	3500	&	LJT+YFOSC(E9G10)	\\
May 20, 2023	&	60084.69	&	1.93	&	3850-9160	&	3500	&	LJT+YFOSC(E9G10)	\\
May 20, 2023	&	60084.72	&	1.96	&	3400-9150	&	350	&	LJT+YFOSC(G3)	\\
May 20, 2023	&	60084.78	&	2.02	&	3850-9160	&	3500	&	LJT+YFOSC(E9G10)	\\
May 21, 2023	&	60085.71	&	2.95	&	3850-9160	&	3500	&	LJT+YFOSC(E9G10)	\\
May 21, 2023	&	60085.74	&	2.98	&	3400-9150	&	350	&	LJT+YFOSC(G3)	\\
May 22, 2023	&	60086.56	&	3.80	&	3900-8870	&	350	&	XLT+BFOSC(G4)	\\
May 22, 2023	&	60086.77	&	4.01	&	3500-9500	&	32000	&	LJT+HRS	\\
May 23, 2023	&	60087.54	&	4.78	&	4000-8000	&	2000	&	XLT+BFOSC(E9G10)	\\
May 23, 2023	&	60087.71	&	4.95	&	3850-9160	&	3500	&	LJT+YFOSC(E9G10)	\\
May 23, 2023	&	60087.74	&	4.98	&	3400-9150	&	350	&	LJT+YFOSC(G3)	\\
May 25, 2023	&	60089.52	&	6.76	&	3900-8870	&	350	&	XLT+BFOSC(G4)	\\
May 26, 2023	&	60090.74	&	7.98	&	3400-9150	&	350	&	LJT+YFOSC(G3)	\\
May 27, 2023	&	60091.69	&	8.93	&	3850-9160	&	3500	&	LJT+YFOSC(E9G10)	\\
May 29, 2023	&	60093.69	&	10.93	&	3400-9150	&	350	&	LJT+YFOSC(G3)	\\
May 31, 2023	&	60095.67	&	12.91	&	3400-9150	&	350	&	LJT+YFOSC(G3)	\\
June 1, 2023	&	60096.71	&	13.95	&	3850-9160	&	3500	&	LJT+YFOSC(E9G10)	\\
June 2, 2023	&	60097.70	&	14.94	&	3400-9150	&	350	&	LJT+YFOSC(G3)	\\
June 4, 2023	&	60099.75	&	16.99	&	3500-9500	&	32000	&	LJT+HRS	\\
June 14, 2023	&	60109.54	&	26.78	&	3900-8870	&	350	&	XLT+BFOSC(G4)	\\
June 17, 2023	&	60112.53	&	29.77	&	3900-8870	&	350	&	XLT+BFOSC(G4)	\\
\hline
\hline
\hline
\end{tabular}

$^a${The epoch is relative to the explosion date, MJD = 60082.76.}

\label{Tab:Spec_log}
\end{table*}

\renewcommand{\thetable}{S\arabic{table}}
\begin{table*}
\caption{Results of H$\alpha$ emission}
\scriptsize
\begin{tabular}{ccccccccc}
\hline\hline
 Phase $^a$ & $\rm L_{H\alpha}$$^b$ & $\rm L^{N}_{H\alpha}$ $^c$  & FWHM$^d$  & $t$$^e$  & radius & $\rho $  &$\rm \dot{M}$ & $\beta^f$\\
 (d) & ($10^{39}\rm erg/s$) &($10^{38}\rm erg/s$)  &(km/s)&  (K) &  (10$^{14}$cm) & $(\rm 10^{-15} g/cm^3)$ &  $(10^{-3} {\rm M}_{\odot}/{\rm a})$ & \\
\hline
	1.79	&	2.71& 3.1	&	55	&	18800	&	2.33	&	7.41	&	0.56	&	1.97	\\
	1.93	&	3.03& 2.8	&	62	&	19100	&	2.59	&	7.26	&	0.68	&	1.67	\\
	2.02	&	3.32&2.7&	79	&	19300	&	2.74	&	7.31	&	0.77	&	1.52	\\
	2.95	&	4.00&2.7&	141	&	20100	&	3.96	&	6.60	&	1.44	&	0.75	\\
	4.95	&	2.45& ... &	867	&	16800	&	5.75	&	5.67	&	2.61	&	0.20	\\
\hline
\hline
\hline
\end{tabular}

$^a${The epoch is relative to the explosion date, MJD = 60082.76.}\\
$^b${Luminosity of narrow and broad components of H$\alpha$.}\\
$^c${Luminosity of the narrow component of H$\alpha$.}\\
$^d${FWHM of narrow H$\alpha$ removed the instrumental affection.}\\
$^e${Photospheric temperature derived from black body fit in Fig. \ref{<bolo>}.}\\
$^f${$\beta\ge 1$ is required to produce a reliable mass-loss rate. }

\label{Tab:ha}
\end{table*}

\setcounter{figure}{0} 
\renewcommand{\thefigure}{S\arabic{figure}}

\begin{figure}
\centering
\includegraphics[width=12cm,angle=0]{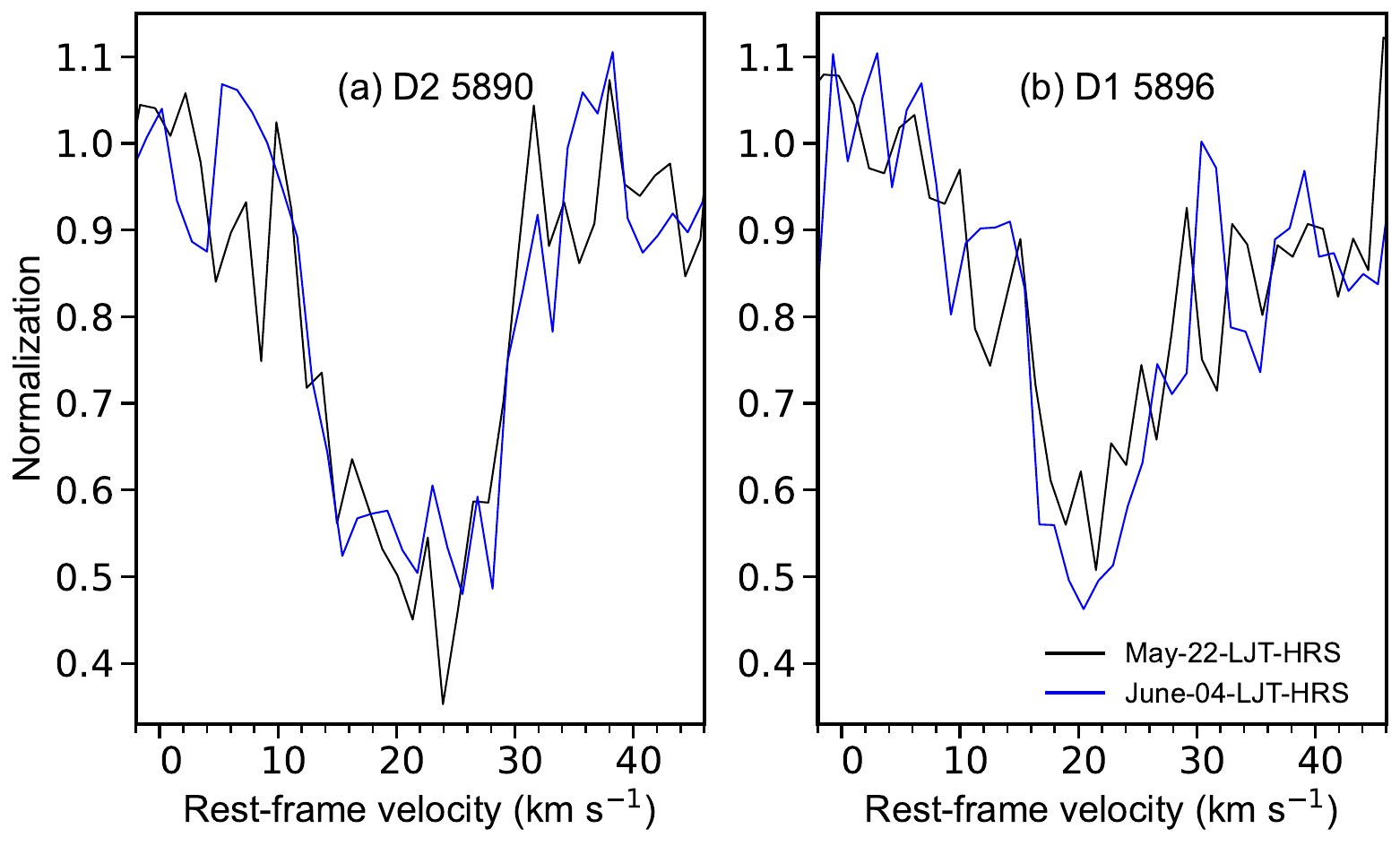}
 \caption{High-resolution spectra of SN 2023ixf around the Na~{\sc i} D absorption in velocity frame.  }
\label{<spe_NaID>}
\end{figure}

\begin{figure*}
\centering
\includegraphics[width=12cm,angle=0]{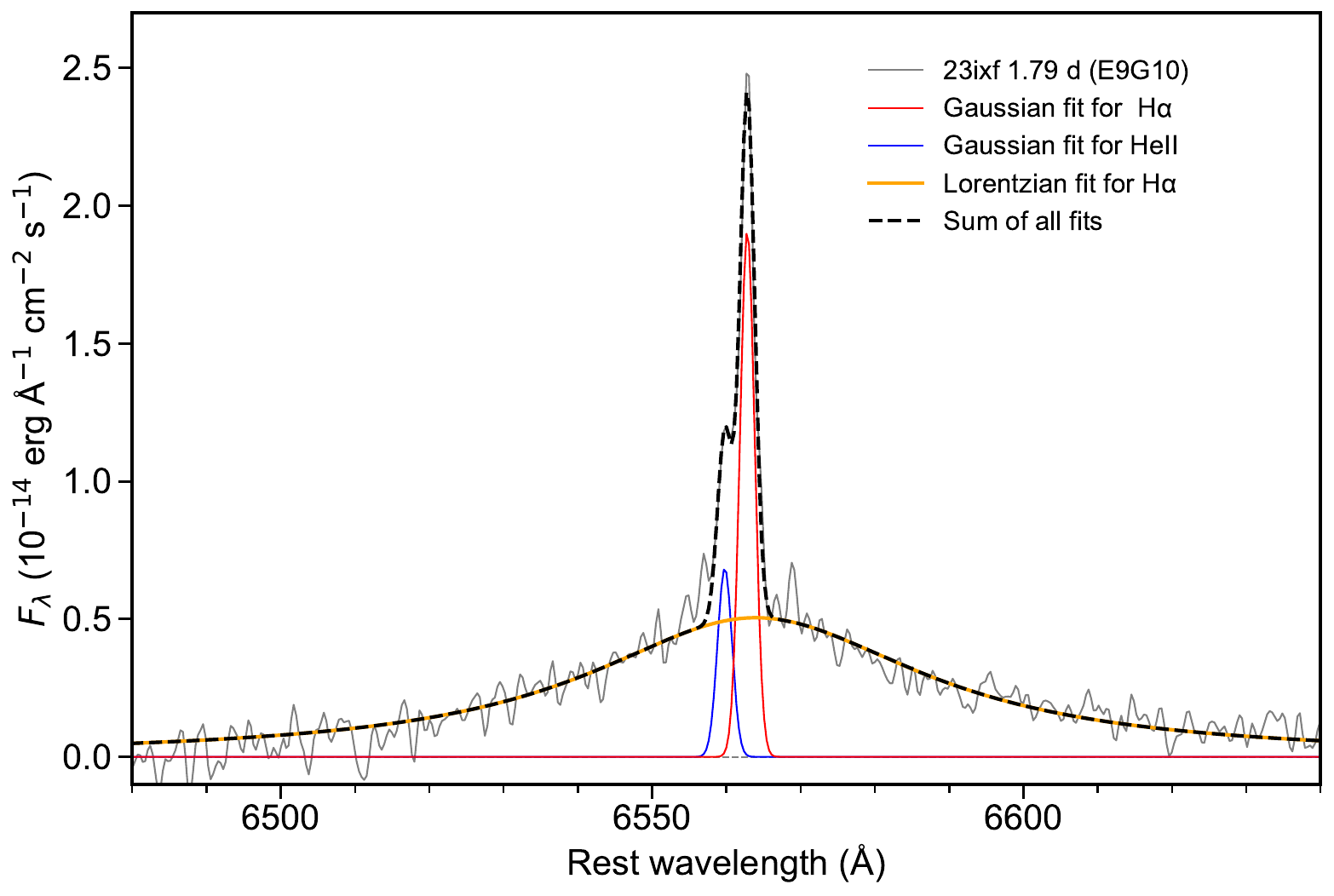}
 \caption{Gaussian and Lorentzian fit the narrow and broad components of the H$\alpha$ line in the YFOSC E9G10 spectrum obtained at 1.79 d after the explosion. A narrow emission at the left side of H$\alpha$ is supposed to be He II 6560.  }
\label{<ha_3fit>}
\end{figure*}

\begin{figure*}
\centering
\includegraphics[width=12cm,angle=0]{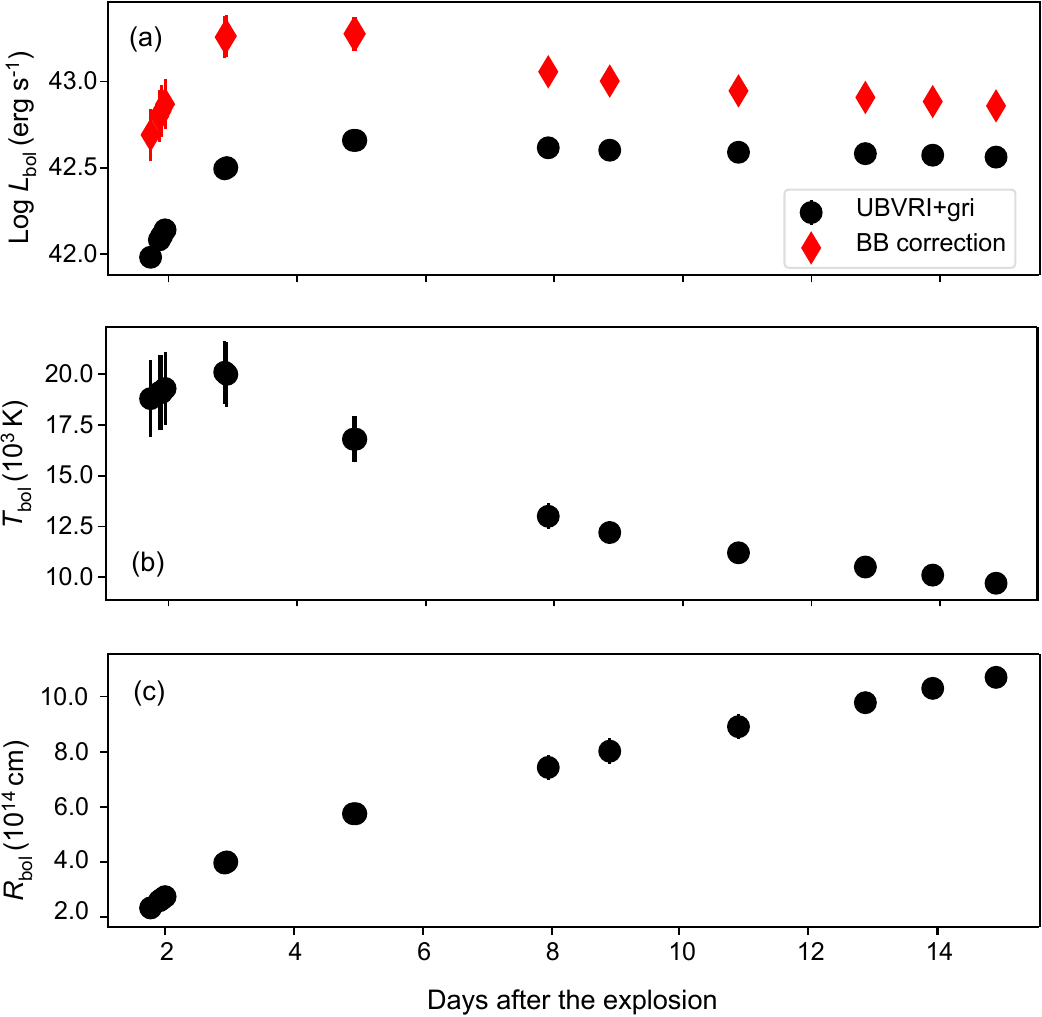}
\caption{Bolometric light curve, black body temperature, and photospheric radius of SN 2023ixf calculated via the photometry obtained at Li-Jiang Observatory and Xing-Long Observatory. (a): Bolometric light curves derived from $UBVRI+gri$-band photometry and black body fit. (b): The temperature derived from the black-body fit of photometry. (c): Photospheric radius derived from luminosity and temperature.   }
\label{<bolo>}
\end{figure*}

\end{appendices}

\end{document}